\documentclass[aps,groupedaddress,showpacs]{revtex4}
\usepackage{graphicx}
\newcommand{\figwidth}{8cm}
\begin{document}

\title{Raman scattering study of charge ordering in $\beta$-Ca$_{0.33}$V$_2$O$_5$}
\author{Z. V. Popovi\'c \footnote {Present address: Materials Science Institute,
University of Valencia, Poligono La Coma s/n, 46980 Paterna
(Valencia) Spain; e-mail: Zoran.Popovic@uv.es}, M. J.
Konstantinovi\'c and V. V. Moshchalkov}

\affiliation{Laboratorium voor Vaste-Stoffysica en Magnetisme,
Katholieke Universiteit Leuven, Celestijnenlaan 200D, B-3001
Leuven, Belgium }

\author{M. Isobe and Y. Ueda}

\affiliation{Materials Design and Characterization Laboratory,
Institute for Solid State Physics, The University of Tokyo, 5-1-5
Kashiwanoha, Kashiwa, Chiba 277-8581, Japan}

\begin{abstract}

Polarized Raman spectra of the calcium vanadium oxide bronze
$\beta$-Ca$_{0.33}$V$_2$O$_5$ are measured in the temperature
range between 300 K and 20 K. The charge ordering phase transition
at about 150 K is characterized by the appearance of the new
Raman-active modes in the spectra, and by anomalies in the
electronic background scattering. The high temperature Raman
scattering spectra of $\beta$-Ca$_{0.33}$V$_2$O$_5$ are in
apparent resemblance with those of $\alpha'$-NaV$_2$O$_5$, which
suggests that there is a similar charge-phonon dynamics in both
compounds. The study of dynamical properties and the symmetry
analysis of the Raman modes show that in the mixed valence state
of $\beta$-Ca$_{0.33}$V$_2$O$_5$ the electrons are delocalized
into V$_1$-O$_5$-V$_3$ orbitals. We propose that in the charge
ordered state below 150 K the $d$-electrons localize within
V$_1$-V$_3$ ladders, either in "zig-zag" fashion like in
$\alpha'$-NaV$_2$O$_5$ or in the forms of the double chains like
in $\gamma$-LiV$_2$O$_5$.

\end{abstract}

\maketitle

The vanadium bronzes, $\beta(\beta')$-A$_x^{1+(2+)}$V$_2$O$_5$
($x=\frac1{3},\, \frac2{3}$; A= Li, Na, Ag, Ca, Sr, Pb, Cu etc.)
exhibit a variety of phenomena that originate from strong electron
correlations. For example, various charge and spin orderings
\cite{a1,a2,a3,a4}, and even superconductivity
($\beta$-Na$_{0.33}$V$_2$O$_5$  \cite{a5} and
$\beta'$-Cu$_{0.65}$V$_2$O$_5$  \cite{a6}) are found in this class
of materials. The electronic properties are not yet fully
understood due to complex crystal structure of these bronzes. The
$\beta$ compounds with divalent $A$ cations such as Ca$^{2+}$ and
Sr$^{2+}$ were first prepared by Bouloux {\it et al.} \cite{a7}.
The X-ray measurements \cite{a8} revealed the presence of a
superstructure with a lattice modulation vector
$\overrightarrow{q}$=(0, 1/2, 0) at room temperature in
$\beta$-A$^{2+}_{0.33}$V$_2$O$_5$ (A = Ca and Sr). This indicates
the doubling of the unit cell along the {\it b}-axis. Such a
superstructure was also observed in
$\beta$-A$^{1+}_{0.33}$V$_2$O$_5$ (A = Na and Ag) \cite{a8} below
room temperature, originating from the ordering of the A cations
along the {\it b}-axis, that is, formation of the A-chains where
the A cations and vacancies alternate with each other.

The magnetic susceptibility of $\beta$-Ca$_{0.33}$V$_2$O$_5$ vs
temperature, $\chi(T)$, shows a slight jump at around 150 K
\cite{a8}, indicating the existence of a phase transition. Above
the transition temperature the magnetic susceptibility is almost
temperature independent, whereas below the transition temperature
it shows a low-dimensional behavior. The magnetic susceptibility
has a maximum around 50 K, then further decreases by reducing the
temperature, with an upturn at lower temperatures which is caused
by the impurities or defects \cite{a8}. At the transition, the
lattice parameters show a discontinues change, typical for a first
order transition. The NMR study \cite{a9} has revealed that above
the transition temperature there is only one kind of the
electronic state of the V site, whereas below transition
temperature T=T$_{CO}$=149 K one finds two kinds of V sites which
can be identified as magnetic V$^{4+}$ and nonmagnetic V$^{5+}$.
This finding strongly suggests the existence of a CO transition or
the appearance of a charge disproportion at T$_{CO}$.

Here we analyze the CO phase transition in
$\beta$-Ca$_{0.33}$V$_2$O$_{5}$ using the Raman scattering
technique. First, we show that CO in
$\beta$-Ca$_{0.33}$V$_2$O$_{5}$ causes similar changes in the
phonon spectra as those found in $\alpha'$-NaV$_2$O$_{5}$. Second,
we discuss the electron localization effects and possible CO
patterns in this bronze.

\begin{figure}
\includegraphics[width=6cm]{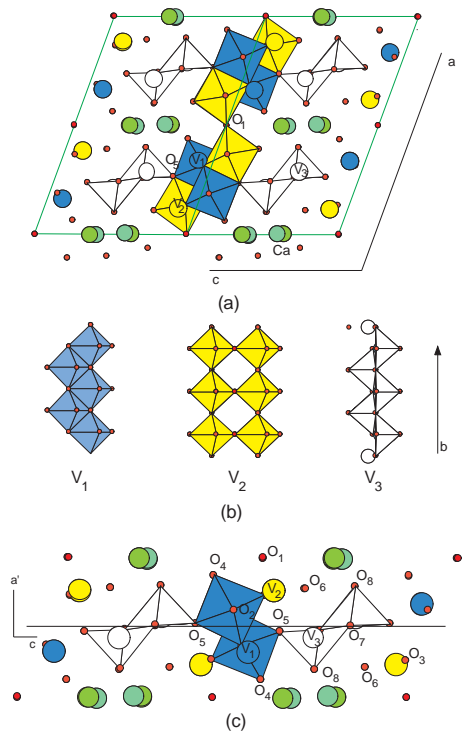}
\caption{\label{fig1}(a) Crystal structure of
$\beta$-Ca$_{0.33}$V$_2$O$_5$ projected on the (ac) plane.  (b)
Three types of the V-chains running parallel to the {\it b}-axis.
(c) Schematic representation of the up-up-down-down orientation of
the VO$_5$ and VO$_6$ polihedra  along the {\it c}-axis }
\end{figure}

Powder samples of $\beta$-Ca$_{0.33}$V$_2$O$_5$ were prepared by a
solid state reaction of mixtures with an appropriate molar ratio
of Ca$_2$V$_2$O$_7$, V$_2$O$_3$ and V$_2$O$_5$. The mixtures were
pressed into pellets and heated at 650$^0$C in an evacuated silica
tube for several days with some intermediate grindings. Details of
the sample preparation were published in Ref. \cite{a8}.

The crystal structure of the $\beta$-phase is monoclinic, space
group $C2/m$ ($C_{2h}^3$) with Z=6 formula units per unit cell. It
has a characteristic V$_2$O$_5$ framework formed by the
edge/corner shearing VO$_5$ and VO$_6$, as shown in Fig. 1(a).
There are three different sites for vanadium atoms: V$_1$, V$_2$
and V$_3$. The V$_2$O$_5$ framework consists of three kinds of
infinite double chains along the {\it b}-axis, as shown in Fig.
1(b). The V$_1$ sites have six-fold, octahedral coordination and
form a zigzag chain by sharing the edges of the VO$_6$ octahedra.
The V$_2$ sites with similar octahedral coordination form a ladder
chain by sharing corners, and the V$_3$ sites, which have 5-fold
square pyramidal coordination, form a zigzag chain by sharing the
edges. The $Ca$ cations are located at the sites in the tunnel
formed by the V$_2$O$_5$ framework. There are two equivalent sites
for $Ca$, but two $Ca$ sites at the same height along the b axis
can not be occupied simultaneously. Therefore, the stoichiometric
composition can be expressed as Ca$_{1/3}$V$_2$O$_5$, that is,
CaV$_6$O$_{15}$.

The Raman spectra were measured from very small crystals of
typical size 5x15$\mu$m embedded in the powder pellets. We used
micro-Raman system with DILOR triple monochromator including
liquid nitrogen cooled charge-coupled-device-detector. The 514.5
nm line of an Ar-ion laser was used as excitation source.
Low-temperature measurements we carried out in the Oxford
continuous flow cryostat with 0.5 mm thick window. The laser beam
was focused by a long distance (10 mm focal length) microscope
objective (magnification 50$\times$).

\begin{figure}
\includegraphics[width=\figwidth]{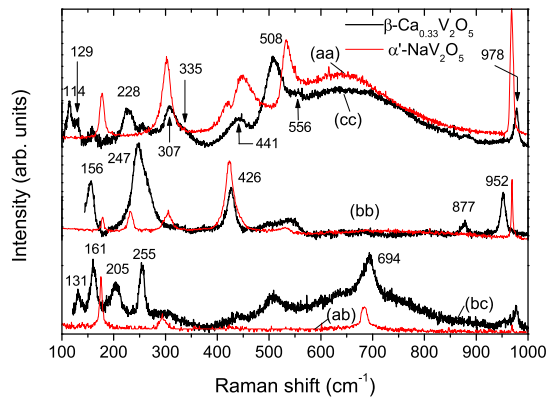}
\caption{\label{fig2} Room temperature polarized Raman spectra of
$\beta$-Ca$_{0.33}$V$_2$O$_5$, together with corresponding Raman
spectra of $\alpha'$-NaV$_2$O$_5$. $\lambda_L$=514.5 nm. }
\end{figure}

The $\beta$-Ca$_{0.33}$V$_2$O$_5$ unit cell consists of six
formula units (Z=6) with 44 atoms in all. Due to that we can
expect a large number of optically active modes. All atoms (except
 O$_1$, which has the $C_{2h}$ site symmetry) have $4(i)$
position symmetry of $C2/m$ $(C_{2h}^3)$ space group \cite {a10}.
Factor-group-analysis (FGA) yields the following distribution of
vibrational modes:

\begin{eqnarray*}
\Gamma= 20A_g(xx, yy, zz, xz)  + 10B_g (xy, yz) + 12A_u ({\bf
E}||{\bf y}) + 24B_u ({\bf E}||{\bf x},{\bf E}||{\bf z}).
\end{eqnarray*}

According to this representation one can expect 30 Raman and 33
infrared active modes (1A$_u$+2B$_u$ are acoustic modes).

The polarized Raman spectra of $\beta$-Ca$_{0.33}$V$_2$O$_5$,
measured from (bc) plane at room temperature, for the parallel and
crossed polarizations, are given in Fig. 2 together with the
polarized Raman spectra of $\alpha'$-NaV$_2$O$_5$. We find that
these spectra coincide with each other in many details, which
suggests structural similarities and similarities in the normal
coordinates of vibrational modes, as we will discuss it later. The
spectra for parallel polarizations consist of the A$_g$ symmetry
modes. Nine modes at 114, 129, 228, 307, 335, 441, 508, 556 and
978 cm$^{-1}$ are clearly seen for the (cc) polarization and five
additional modes at 156, 247, 426, 877, and 952 cm$^{-1}$ for the
(bb) polarization. For the crossed (bc) polarization five Raman
active ${\rm B_{g}}$ symmetry modes at 131, 161, 205, 255 and 694
cm$^{-1}$ are observed. As it is already discussed in the case of
AV$_2$O$_5$ \cite{a11,a12}, the phonon modes in the spectral range
below 500 cm$^{-1}$ originate from the bond bending vibrations,
whereas the higher frequency modes originate from the stretching
vibrations of the V-O ions. The highest frequency mode at 978
cm$^{-1}$ represents V$_3$-O$_8$ (see Fig. 1) stretching
vibrations (V-O$_1$ bond stretching vibration in sodium vanadate
\cite{a11}), whereas the modes at about 877 and 952 cm$^{-1}$
originate from V$_2$-O$_6$ and V$_1$-O$_4$ bond stretching
vibrations, respectively. These modes appear at highest
frequencies, because they are coming from the shortest distance
V-O bonds (V$_3$-O$_8$ in the VO$_5$ pyramid, V$_2$-O$_6$ and
V$_1$-O$_4$ bonds in the VO$_6$ octahedra, see Fig. 1). Note that
the V-O bond stretching vibrational modes in the octahedral
surrounding have lower frequencies than in the pyramidal one, as
it is already found in Ref. \cite {a13}.

\begin{figure}
\includegraphics[width=\figwidth]{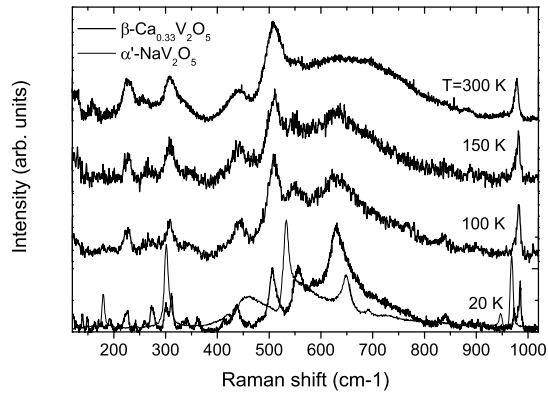}
\caption{\label{fig3} The (cc) polarized Raman spectra of
$\beta$-Ca$_{0.33}$V$_2$O$_5$ at different temperatures. Thin line
is the (aa) polarized spectra of $\alpha'$-NaV$_2$O$_5$ at T=20
K.}
\end{figure}

B$_{1g}$ symmetry mode at 694 cm$^{-1}$ corresponds to the bond
stretching vibrations of V$_3$ and O$_{7}$ ions along the {\it
b}-axis. Since the V-O$_{2}$ distance in
$\beta$-Ca$_{0.33}$V$_2$O$_5$ is close to the same distance
(V-O$_{2b}$) in $\alpha'$-NaV$_2$O$_5$ there is no significant
mode frequency difference between these two oxides. The next
A$_{1g}$ mode of $\beta$-Ca$_{0.33}$V$_2$O$_5$ appears at
frequency of 508 cm$^{-1}$. The frequency of the analogous mode in
$\alpha'$-NaV$_2$O$_5$ is 534 cm$^{-1}$. This mode represents the
bond stretching vibration of V$_3$ and O$_{7}$ ions, along the
{\it c}-axis. Frequency difference between these modes is in
accordance with the difference between values of the
V$_3$-O$_{7c}$ bond lengths in $\beta$-Ca$_{0.33}$V$_2$O$_5$ (2.0
$\AA$) and V-O$_{2a}$ in $\alpha'$-NaV$_2$O$_5$ (1.985 $\AA$).

The Raman mode at 448 cm$^{-1}$ in $\alpha'$-NaV$_2$O$_5$ is
related to the V-O$_3$-V bending vibrations \cite{a11}(mostly
vibration of oxygen ions in the rungs of the ladder structure of
sodium vanadate along the {\it c}-axis direction). As we discussed
in \cite{a11,a12} the force constant of this mode is strongly
affected by the charge of the electrons in the rung. Because of
that, this mode appears at frequency below its intrinsic value of
485 cm$^{-1}$ in pure V$_2$O$_5$ or 470 cm$^{-1}$ in CaV$_2$O$_5$
(where spin electrons are attached to the V$^{4+}$ ions)
\cite{a11}. In addition, this mode in $\alpha'$-NaV$_2$O$_5$ shows
strong asymmetry and broadening. Since this mode has the same
shape and position in both $\alpha'$-NaV$_2$O$_5$ and
$\beta$-Ca$_{0.33}$V$_2$O$_5$ we conclude that the electrons in
$\beta$-Ca vanadate bronze are also delocalized in the V-O-V
orbitals , where the O denotes common corner oxygen ions between
the VO$_{5(6)}$ polihedra. There are two bridge oxygens in
different VO$_{5(6)}$ polihedra: O$_1$ and O$_5$ (Fig.1). Since
O$_1$ is in the center of inversion, its vibration does not
contribute to the Raman scattering process. Thus, only the O$_5$
ion vibration can be included in the normal coordinate of the 441
cm$^{-1}$ V-O-V mode. Furthermore, there are three possible
V-O$_5$-V bonds for the normal coordinate of the 441 cm$^{-1}$
mode: V$_2$-O$_5$-V$_1$, V$_2$-O$_5$-V$_3$ and V$_1$-O$_5$-V$_3$.
Since only the V$_1$-O$_5$-V$_3$ bond has nearly the same bond
length (3.74 $\AA$) and bond angle (136$^o$) as in
$\alpha'$-NaV$_2$O$_5$ \cite{a11}, we conclude that 441 cm$^{-1}$
mode corresponds to the V$_1$-O$_5$-V$_3$ bond bending vibration.
This means that although V$_1$ and V$_3$ are different site V
ions, they are in the mixed valence state (V$^{4.5+}$), whereas
V$_2$ are all in the 5+ states at room temperature. The
crystalographic similarity between $\alpha'$-NaV$_2$O$_5$ and
$\beta$-Ca$_{0.33}$V$_2$O$_5$ is illustrated in Fig.1(c). Note
that the VO$_5$ pyramids and VO$_6$ octahedra are oriented along
the {\it c}-axis as "up-up-down-down" pattern in
$\beta$-Ca$_{0.33}$V$_2$O$_5$ similar to the VO$_5$ pyramidal
structure along the {\it a}-axis in $\alpha'$-NaV$_2$O$_5$.

Fig. 3 shows the Raman spectra of $\beta$-Ca$_{0.33}$V$_2$O$_5$
for the (cc) polarization at different temperatures. By lowering
the temperature all observed modes harden and two new modes at 555
and 630 cm$^{-1}$ appear from the broad structure peaked at about
650 cm$^{-1}$. These two modes become more pronounced below the
transition temperature (150 K), and they are the most intense ones
at the lowest temperature of about 20 K we used in our
experiments. At temperatures less than 100 K almost all modes
split as a consequence of charge ordering followed by the doubling
of the unit cell. This causes the appearance of the zone boundary
modes in the Raman spectra.

In the charge ordered phase there are two possible ways to
redistribute the $d$-electrons from the V$_1$-O$_5$-V$_3$
orbitals:

{\it (i)} Below T$_{CO}$, the spin electrons from the
V$_1$-O$_5$-V$_3$ orbitals localize at the V$_1$ and V$_3$ sites,
forming zigzag ordering pattern similar to the low-temperature
phase of $\alpha'$-NaV$_2$O$_{5}$, Fig.4(a).

{\it (ii)} The spin electrons from the V$_1$-O$_5$-V$_3$ orbitals
localize at the V$_1$ ions forming a 1-D magnetic double-chains
(Fig.4(b)), similar to those in the LiV$_2$O$_5$ \cite{a14}.

\begin{figure}
\includegraphics[width=8cm]{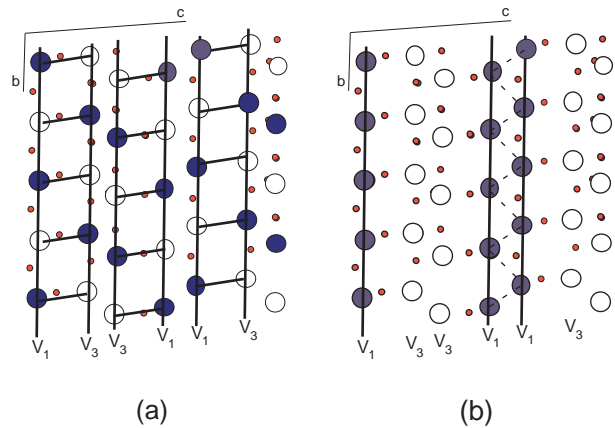}
\caption{\label{fig4} Possible charge ordering patterns in
$\beta$-Ca$_{0.33}$V$_2$O$_5$. (a) Asymmetric spin ladder
structure with a zigzag charge ordering. (b) 1D-double V$_1$
zigzag chain (zigzag ladder) charge ordering.}
\end{figure}

The Madelung energy calculation \cite{a15} of
$\beta$-A$_{0.33}$V$_2$O$_5$ revealed that the lowest-energy
configuration of the $d$-electrons is to form the 1D V$_1$-chains
at half filling when A are divalent cations. However, from the
comparison between the low temperature Raman spectra of
$\beta$-Ca$_{0.33}$V$_2$O$_5$, $\alpha'$-NaV$_2$O$_5$, Fig.2, and
LiV$_2$O$_5$ (Fig. 3 in Ref.\cite{a14}), it is difficult to
unambiguously conclude which CO pattern is realized in the
low-temperature phase of $\beta$-Ca$_{0.33}$V$_2$O$_5$. The
difficulty arises because of the lack of complete knowledge of
phonon dispersions (lack of an appropriate single crystal). As it
is discussed in \cite{a16}, from the measurements of all
inequivalent polarized Raman and infrared spectra it is possible
to exclude some of the CO patterns. However, even then the
additional problems, that do not relieve the assignment, may come
from the strong resonant effects \cite{a17}. Thus, we can suggests
the optical measurements (ellipsometry) as an appropriate method
which can distinguish between these two CO configurations. Namely,
the presence (or absence) of the $\sim$1.1 eV peak \cite{a18} (see
also Fig. 3 in Ref.\cite{a19}) in the absorption spectra of
$\beta$-Ca$_{0.33}$V$_2$O$_5$ along the {\it b}-axis, should
confirm the "zigzag" (or double chain) charge ordering pattern. At
the moment such measurements are not possible because of lack of
single crystal samples.

In conclusion, the Raman scattering spectra of
$\beta$-Ca$_{0.33}$V$_2$O$_5$ above and below the phase transition
temperature of about 150 K, show similar changes in the phononic
and electronic excitations as in $\alpha'$-NaV$_2$O$_5$. At
temperatures above the phase transition temperature, the
broadening and the asymmetry of the 441 cm$^{-1}$ mode suggest
that the electrons are delocalized into the V$_1$-O$_5$-V$_3$
orbitals. Below the phase transition, the charge ordering takes
place, and the $d$-electrons order either in the "zigzag" fashion
along the V$_1$-V$_3$ ladders, or at double chain V$_1$ ions.

\section*{Acknowledgment}
Z.V.P. and M.J.K acknowledges support from the Research Council of
the K.U. Leuven and DWTC.  The work at the K.U. Leuven is
supported by the Belgian IUAP and Flemish FWO and GOA Programs.

\end{document}